\documentclass[aps,pra,onecolumn,preprint]{revtex4-1}
\usepackage{amsmath}
\usepackage{epsfig}
\usepackage{graphicx}
\usepackage{color,float}
\usepackage{amssymb}
\usepackage{epstopdf}
\usepackage[colorlinks=True, linkcolor=blue, citecolor=blue]{hyperref}
\usepackage{xcolor}
\begin{document}
\title{Fixed-Point Few-Body Hamiltonians in Quantum Mechanics}
\thanks{
Dedicated to Steven Weinberg's memory, celebrating  his contributions on chiral nuclear forces. 
}     
\author{Lauro Tomio}
\affiliation{Instituto de F\'isica Te\'orica, Universidade Estadual Paulista - UNESP, 01156-970, S\~ao Paulo, SP, Brazil}
\author{T. Frederico}
\affiliation{Instituto Tecnol\'ogico de Aeron\'autica, DCTA, 12228-900, S\~ao Jos\'e dos Campos, SP, Brazil}
\author{V. S. Tim\'oteo}
\affiliation{Grupo de Optica e Modelagem Num\'erica, Faculdade de Tecnologia, 
GOMNI/FT - Universidade Estadual de Campinas, 13484-332, Limeira, SP, Brazil.}  
\author{M. T. Yamashita}
\affiliation{Instituto de F\'isica Te\'orica, Universidade Estadual Paulista - UNESP, 01156-970 S\~ao Paulo, SP, Brazil}  
\begin{abstract}
We revisited how  Weinberg's ideas in Nuclear Physics influenced
our own work and lead to a renormalization group invariant framework 
within the quantum mechanical few-body problem, and we also update  
the discussion on the relevant scales in the limit of short-range 
interactions. In this context, it is revised the formulation of the 
subtracted scattering equations and fixed-point Hamiltonians applied 
to few-body systems, in which the original interaction contains 
point-like singularities, such as Dirac-delta and/or its derivatives.
The approach is being illustrated by considering two-nucleons 
described by singular interactions.
This revision also includes an extension of the renormalization
formalism to three-body systems, which is followed by an updated 
discussion on the applications to four particles.
\end{abstract}
\maketitle

\section{Introduction}~\label{intro}

The use of effective interactions containing singularities at short distances has been motivated 
in nuclear physics 
by the development of a chirally symmetric nucleon-nucleon interaction, which contains contact
interactions, as represented by the Dirac-delta function and its higher order derivatives. 
A first approach in this direction, following a review on phenomenological
Lagrangian~\cite{1979wein}, was established by Steven Weinberg, when describing nuclear 
forces and nucleon-nucleon interactions derived from effective 
chiral Lagrangians~\cite{1990wein,1991wein,1992wein}.  
Therefore, the main motivation in considering point-like interactions emerged due to applications 
of effective theories to represent a more fundamental theory, supposed to be the 
Quantum ChromoDynamics (QCD), which has been found too complex to be accessible by exact 
approaches. Of particular interest is the fact that such effective theories allow one to 
parametrize the physics of the high momentum states and work with effective degrees of freedom. 

The association of limit cycles and fixed point Hamiltonians to renormalization group methods 
have being known for a long time, since some original studies led by Kenneth Wilson 
on renormalization group applied to 
strong interactions~\cite{1970wilson,1971wilson,1974wilson,1983wilson}.  
About two decades after the first studies on that, these investigations start to be more 
effectively explored by  Wilson itself together with other collaborators in a series of 
papers, considering general studies of renormalization of
Hamiltonians~\cite{1993glazek,1993glazek2,1994glazek,1997perry,1998glazek}. 

The idea to use an effective renormalized Hamiltonian, which  includes the coupling between low- 
and high-momentum states was suggested in Refs.~\cite{1993glazek,1993glazek2,1994glazek}, 
 with the renormalized Hamiltonian carrying the physical information contained in the quantum 
 system at high momentum states.
With particular applications to QCD and other field theories on the light cone,
an appropriate detailed review considering non-perturbative renormalization can be found 
in Ref.~\cite{1998brodsky}.
As concerned with non-relativistic quantum mechanics, Ref.~\cite{1997lepage} is also providing 
some clear examples on the application of renormalization ideas. 

From the renormalization group approach, the concept of universality and limit-cycle behavior 
were further explored by  Wilson and Glazek in Ref.~\cite{2004glazek}, within a simple 
Hamiltonian model, analytically soluble at criticality, which exhibits an infinite exact 
geometric series of  bound-state energy eigenvalues.
Within nuclear physics studies considering three-nucleons and  halo-nuclei systems,  
universal aspects of renormalized three-body systems have been established at that time in 
Refs.~\cite{1992amorim,1997amorim}. These studies with renormalized singular contact interactions 
were also followed by establishing  correlations between low-energy observables of 
three-atom systems, with the emergence of scaling limits in weakly-bound triatomic systems  
in Ref.~\cite{1999frederico} (see, also Refs.~\cite{1999tomio,2000delfino,2000delf}). 

By following an application considering the renormalization of the one-pion-exchange 
potential  plus a Dirac-delta interaction to nuclear physics~\cite{1990wein,1991wein,1992wein},
reported in Ref.~\cite{1999fred}, it was proposed in Ref.~\cite{2000frederico}
a general non-perturbative renormalization scheme to treat singular interactions in 
quantum mechanics, considering a subtraction procedure in the propagator within the kernel 
of the scattering equation. The procedure 
was based in the renormalization group invariance of quantum mechanics.
Such renormalization approach has been applied successfully to several works related to the 
renormalization of chiral nuclear forces, by using multiple subtractions~\cite{2005tim,2007tim,2007timNPA,2011tim,2011std,2012szpigel}. 
In the context of nuclear physics, the approach has been discussed in more recent reviews, 
as Refs.~\cite{2011birse,2012fred,2017batista,2020ham,2020epel,2021batista,2021entem,2021timoteo}. 
In an effective QCD-inspired theory of mesons the method was also applied 
in Ref.~\cite{2001frederico}. It generalizes some ideas suggested in
Refs.~\cite{1995Adhikari1,1995Adhikari2}, by performing $n$ subtractions in the free propagator 
of the singular scattering equation at an arbitrary energy scale, where $n$ is the smallest 
number necessary to regularize the integral equation. The unknown short range physics related 
to the divergent part of the interaction are replaced by the renormalized strengths of the 
interaction, which are known from the scattering amplitude at some reference energy. 
In this context, the renormalization scale is given by an arbitrary subtraction point, with the  
existence of a sensible theory for singular interactions relying on the property that the
subtraction point can slide without affecting the physics of the renormalized theory~\cite{wein1}. 

Within the proposal in Ref.~\cite{2000frederico}, for the renormalization group invariance of
quantum mechanics, a subtraction point is a defined scale at which the quantum mechanical 
scattering amplitude is known. A fixed-point 
Hamiltonian~\cite{1970wilson,1974wilson,1983wilson,1998fisher,2002zinnjustin} should 
have the property to be stationary in the parametric space of Hamiltonians, as a function of the 
subtraction point~\cite{2002zinnjustin}.  For the realization of this property, it is required 
that the derivative of the renormalized Hamiltonian in respect to the renormalization scale 
is zero. This implies that the scattering amplitude does not depend on the arbitrary 
subtraction scale, and in the corresponding renormalization group equations. 
As shown in ref.~\cite{2000frederico}, due to the requirement that the physics of the theory 
remains unchanged, the driving term of the $n-$subtracted scattering equation changes as 
the subtraction point moves. Such driving term satisfies the quantum mechanical 
Callan-Symanzik (CS) equation~\cite{1970Callan,1970Symanzik1,1970Symanzik2}, 
which is a first order differential equation with respect to the renormalization scale. 
As verified, the renormalization group equation (RGE) matches the quantum mechanical 
theory at scales $\mu$ and $\mu+d\mu$, without changing its physical content~\cite{1993georgi}.

The purpose of this contribution is to present a brief review on the impact of the concepts 
put forward by Weinberg in our own work. We revisit our studies on the 
renormalization group invariance approach and on the fixed-point 
(renormalized) Hamiltonian for a quantum mechanical few-body system, in consistency 
with Ref.~\cite{2000frederico}.
Here, we are partially following a previous unpublished work by some of us, available in
Ref.~\cite{preprint}, where the general concept of fixed-point Hamiltonians is unified to 
the practical and useful theory of renormalized scattering equations~\cite{2000frederico}. 
We provide response to a question, which is relevant from the theoretical and practical points 
of view, on the existence and formulation of the corresponding renormalized Hamiltonian of a 
quantum mechanical few-body systems (which could be further generalized), when the original 
interaction contains singular terms.  By working in the momentum space, we 
illustrate the method by diagonalizing a renormalized Hamiltonian through an example where up 
to three bound-state energies are shown to converge to the same exact results (in the limit of 
infinite momentum cut-off), irrespectively to the value of the energy-scale parameter being used 
for the renormalized theory.  We also review  a detailed example on how to construct a 
fixed-point Hamiltonian in a situation where higher singularities are describing the 
original interaction. In the last example, we review the application of the subtracted 
renormalizaton scheme for the one-pion-exchange potential plus a Dirac-delta interaction 
based directly in the Weinberg ideas applied to the nucleon-nucleon scattering. We also 
review the extension of the method  of subtracted equations and the renormalized Hamiltonian 
to three-particles and also beyond that.

The next sections of the present contribution are organized as follows:    
The formalism for a fixed-point Hamiltonian is detailed in the next section~\ref{sec2}. 
In section~\ref{sec3}, we workout a few examples of the subtraction approach applied to 
renormalize two-body Hamiltonians with original singular interactions.
In this section we also show how to construct a fixed-point Hamiltonian for the case 
that higher order singularities exist in the original interaction.
In section~\ref{sec4}, by considering three-body systems, the subtracted renormalization 
approach is applied to the Faddeev formalism. 
The case of four particles, for which the approach requires a new scale, 
is shortly discussed in section~\ref{sec5}.
Finally, in section~\ref{sec6}, we have our concluding remarks. 

\section{Renormalized Hamiltonian}\label{sec2}

In this section, by assuming an effective interaction $V_{\cal R}$, with a free Hamiltonian
$H_0$, we introduce the renormalized Hamiltonian, which is a fixed point operator, namely, 
it is independent on the subtraction point, given by
\begin{equation}
H_{\cal R} = H_0 + V_{\cal R},
\label{1}\end{equation}
The two-body Lippmann-Schwinger (LS) equation for the scattering T-matrix, obtained 
from the renormalized  Hamiltonian $H_{\cal R}$, for the free Green's function propagator, 
forward in time, $G^{(+)}_0(E)=(E+{\rm i}\epsilon -H_0)^{-1}$, where $E$ is the total energy, 
can be written as
\begin{eqnarray}
T_{\cal R}(E)&=&V_{{\cal R}}+V_{{\cal R}}~G^{(+)}_0(E)~T_{\cal R}(E) 
= V_{{\cal R}}+V_{{\cal R}}~(E+{\rm i}\epsilon -H_0)^{-1}~T_{\cal R}(E)\ ,
\label{2}
\end{eqnarray}
where the label ${\cal R}$ indicates that the T-matrix is the renormalized one, which is finite and
containing the physical information to fix it. This T-matrix, obtained from the effective interaction 
$V_{{\cal R}}$, by following the renormalization procedure, contains all the necessary counter-terms to subtract the infinities originated by iterations of the LS equation.
Therefore, one  should be able to derive the $n-$subtracted T-matrix equation~\cite{2000frederico}
from  the LS equation with the renormalized potential  $V_{\cal R}$,  which should also provide the
perturbative renormalization of the T-matrix. Furthermore, $V_{\cal R}$ should lead to the CS equation for the evolution of the driving term  of the subtracted scattering equation  with the subtraction point, which is arbitrary, namely the ``sliding scale" a concept clearly explained in Weinberg book on Quantum Field Theory~\cite{wein1} and perfectly adaptable to quantum mechanics.

Within the renormalization approach to the LS equation with singular interactions, Dirac-delta and 
its derivatives, given in~\cite{2000frederico}, the fixed-point interaction $V_{\cal R}$ is identified with the driving term derived from the $n-$th order subtracted $T$-matrix equation, when it is rewritten in the standard form of the LS equation,  as given by Eq.~(\ref{2}).

The driving term of the subtracted $T-$matrix equation~\cite{2000frederico} is denoted by
$V^{(n)}(-\mu^2;E)$, where $-\mu^2$ is the subtraction point, that for convenience is 
chosen to be negative value of energy. Note the dependence on the energy $E$, and  
$n$ is the number (order) of subtractions necessary to turn finite  the solution of the LS 
equation, providing an enough number of subtractions to have the integral equation regularized. 
The $n-$th order subtracted LS equation for
the T-matrix is written as~\cite{2000frederico}:
\begin{equation}\label{3}
T(E)=V^{(n)} (-\mu^2;E)
+ V^{(n)}(-\mu^2;E)~G^{(+)}_n(E;-\mu^2)~T(E)
\end{equation}
where the driving term  
$V^{(n)}\equiv V^{(n)}(-\mu^2;E)$ is built recursively:
\begin{equation}
V^{(n)}\equiv \left[1-(-\mu^2-E)^{n-1}V^{(n-1)}
~G^{n}_0(-\mu^2)\right]^{-1}~V^{(n-1)}
+V_{sing}^{(n)}(-\mu^2) , \label{4}
\end{equation}
and the $n$-th subtracted Green's function is
$G^{(+)}_n(E;-\mu^2)\equiv\left[(-\mu^2-E)G_0(-\mu^2)\right]^n 
G^{(+)}_0(E).$ 

The renormalization constants, brought with he higher-order singularities of the two-body 
potential,  are introduced 
in $V^{(n)}$ through $V_{sing}^{(n)}(-\mu^2)$, which are determined by physical
observables. We observe that for the studies we have performed  with the subtracted LS 
equations considering singular interactions for the NN scattering~\cite{1999fred,2005tim,2011tim}, 
the partial wave S-matrix resulted unitary.

The fixed-point interaction $V_{\cal R}$ is derived from Eq.~(\ref{3}) by rearranging 
terms, adding and subtracting $V^{(n)}G^{(+)}_0(E) T(E)$ and demanding that  
$T_{\cal R}(E) = T(E)$, with $T(E)$ satisfying the standard LS equation, which results in:
\begin{eqnarray}
V_{{\cal R}} =\left[{1+V^{(n)}\left(G^{(+)}_0(E)-
G^{(+)}_n(E;-\mu^2)\right)}\right]^{-1} 
V^{(n)} \, .\label{7}
\end{eqnarray}
The renormalized interaction by itself is not well defined for singular interactions; 
nevertheless, the T-matrix solution of the standard LS equation (\ref{2}) is finite, due 
to the obvious equivalence with the $n-$th order subtracted equation for the T-matrix. 
Essentially, the subtractive renormalization procedure for the T-matrix equation was 
instrumental to write the renormalized fixed-point interaction given in Eq.~(\ref{7}).
In what follows, it should be understood that the T-matrix refers to the renormalized one
($T=T_{\cal R}$), such that we drop the index $\cal R$ from it.
However, $V^{(n)}$ should be distinguished from $V_{\cal R}$.

In  practical applications using $V_{{\cal R}}$ to get the eigenvalues and eigenstates of the 
renormized Hamiltonian,
 an ultraviolet  momentum cut-off  ($\Lambda $) has to be introduced in the calculation. 
 The limit $\Lambda\rightarrow \infty$ can be approached numerically for large values of 
 the cut-off and the results 
should be the same as the ones obtained through the direct use of the subtracted LS equations, 
as in the  case of the bound state eigenvalues of the renormalized 
Hamiltonian. This behavior will be illustrated in subsection~\ref{subsec3.1}, where we 
use a Dirac-delta plus a Yukawa potential and we compute by diagonalization of the 
renormalized Hamiltonian several bound state energies. 

We observe that the physical inputs associated with the singular part of the interaction are 
introduced through $V^{(n)}$ that contains the renormalized coupling constants given at some 
energy scale $-{\mu}^2$, which are introduced in a recursive from.
In the case of a potential that includes a Dirac-delta, one subtraction in the kernel of the 
integral equation is enough to obtain meaningful physical results
from the solution of the subtracted T-matrix  equation. Going to higher singular potential, 
like for example, the Laplacian of the
Dirac-delta,  at least three subtractions are necessary to turn finite the 
T-matrix~\cite{2000frederico}.
For a short-range non-singular potential $V$, we demand $V_{sing}^{(n)}=0$ 
and obviously $V_{{\cal R}}= V $, as the renormalized T-matrix is indeed the one obtained 
from the standard LS equation.

The subtraction point in the renormalized interaction is arbitrary, and should not 
change the physical content
of the model. The renormalization group method allows to arbitrarily change this 
prescription, implied by the independence of $V_{{\cal R}}$ on the subtraction 
point, which maintains invariant the associated physics. 
From that, a definite prescription to modify $V^{(n)}$ in Eq.~(\ref{3}) can be 
derived, preserving the model outcomes. Therefore, $V_{\cal R}$ with the
associated T-matrix [as given by Eq.~(\ref{2})], and $H_{\cal R}$ are 
independent on $\mu^2$,
\begin{eqnarray}
\frac{\partial V_{\cal R}}{\partial \mu^2}= 0, 
\;\;\; \frac{\partial T(E)}{\partial \mu^2} =0\;\;\;
{\rm and}\;\;\;
\frac{\partial H_{\cal R}}{\partial \mu^2}= 0 \,,
\label{11}\end{eqnarray}
which means that $H_{\cal R}$ is a fixed-point Hamiltonian independent on the 
subtraction scale. 

The Callan-Symanzik renormalization group equation in quantum mechanics~\cite{2000frederico} 
for the driving term $V^{(n)}$ of the subtracted LS equation for the $T-$matrix follows 
from Eqs.~(\ref{7}) and~(\ref{11}):
\begin{eqnarray}
 \frac{\partial V^{(n)}}{\partial \mu^2} = -V^{(n)}
\frac{\partial G^{(+)}_n(E;-{ \mu}^2) }{\partial \mu^2}
V^{(n)} \, ,
\label{13}
\end{eqnarray}
with the boundary
condition $V^{(n)}= V^{(n)}(-{{\mu }}^{2};E)$ at a 
reference scale ${\mu }$. 
The driving term $V^{(n)}$ solution of the
differential equation (\ref{13}) is equal to
$T(E=-\mu^{2})$, relating the subtraction scale to the energy dependence of the
T-matrix itself. 
In the case of $n=1$, Eq.~(\ref{13}) is a
differential form of the renormalized LS equation for the T-matrix
\begin{equation}
\left. \frac{d}{dE}T(E)\right| _{E=-\mu ^{2}}=-T(-\mu ^{2})G_{0}^{2}(-\mu
^{2})T(-\mu ^{2}).  \label{14}
\end{equation}
To summarize: the fixed-point
Hamiltonian is invariant under renormalization group transformation for
singular potentials.  Such Hamiltonian contains the finite
coefficients/operators with the physical information about the
quantum mechanical system, in addition it includes the necessary counter 
terms that make finite the scattering amplitude. 

The above formalism is illustrated by three examples with the application of the 
renormalization approach by using the subtraction method and the fixed point 
Hamiltonian to two-body problems.
In the first example, we consider the numerical diagonalization of the 
regularized form of the fixed-point Hamiltonian for a Yukawa plus a 
Dirac-delta interaction. By computing the corresponding eigen-energies, 
it was demonstrating that they are independent on the momentum cut-off
driven to infinity. In the second example, 
the explicit form of the renormalized potential is revisited, by 
considering a four-term-singular bare interaction~\cite{2012fred}.
The third example is an application of the subtracted renormalization 
scheme to the two-nucleon system, by using the one-pion-exchange 
potential supplemented by contact interactions~\cite{1999fred}.
Particularly, in this third example, we notice that this 
renormalization procedure is also suitable for the case in which the main part of  
the one-pion-exchange potential is singular at the origin, as it happens with the 
tensor part of the interaction that goes with $r^{-3}$.

\section{Examples of the subtraction method for singular interactions}\label{sec3}
\subsection{One-term singular renormalized Hamiltonian diagonalization}\label{subsec3.1}

Let us consider a two-body interaction composed by a regular Yukawa potential 
plus a singular Dirac-delta interaction, such that the corresponding matrix 
elements in the momentum space are given by
\begin{equation}
\langle \vec{p} |V_{\cal R}| \vec{q} \rangle 
= \langle \vec{p} |V| \vec{q} \rangle +\frac{\lambda_\delta}{2\pi^2}
=-\frac{1}{2\pi^2}
\left(\frac{2}{|\vec{p}-\vec{q}|^2+\eta^2}-\lambda_{\delta}\right),
\label{15}
\end{equation}
where $\lambda_\delta$ is the strength of the singular part of the renormalized or
fixed point interaction, with $\eta$ being a constant given by the 
inverse of the range of the regular part of the interaction.
The strength $\lambda_\delta$ can be derived from the full renormalized 
T-matrix, considering the following operator expression, obtained
from Eq.~(\ref{2}):
\begin{eqnarray}
\left(1-VG^{(+)}_0(E)\right)T_{\cal R}(E)= 
V+ |\chi\rangle\frac{ \lambda_{\delta}}{2\pi^2}\langle\chi| 
\left[1+ G^{(+)}_0(E)T_{\cal R}(E)\right] \ ,
\label{18}
\end{eqnarray}
where $\langle \vec p|\chi\rangle=1$ is the form factor associate with the 
Dirac-delta potential. By defining the  T-matrix of the regular potential $V$ as
{\small$T^V(E)= \left(1-VG^{(+)}_0(E)\right)^{-1}V,$} and
using the identity $\left(1-VG^{(+)}_0(E)\right)^{-1}
= \left(1+T^V(E)G^{(+)}_0(E)\right)$, we obtain
{\small \begin{eqnarray}\hskip -0.7cm
T_{\cal R}(E)&=&T^V(E)+ 
\frac{\left( 1+T^V(E)G^{(+)}_0(E)\right)|\chi\rangle
\langle\chi|\left(1+G^{(+)}_0(E)
T^V(E)\right)}
{\displaystyle \frac{2\pi^2}{\lambda_{\delta}}-
\langle\chi|G^{(+)}_0(E)|\chi\rangle -
\langle\chi|G^{(+)}_0(E)T^V(E)G^{(+)}_0(E)|\chi\rangle} 
\ . \label{19}
\end{eqnarray}
}In this case, one subtraction is enough to render finite the theory.
Therefore, at the subtraction point $-\mu^2$, 
the above defines the T-matrix of Eq. (\ref{4}) for $n=1$:
{\small\begin{eqnarray}
V^{(1)}&=&T^V(-\mu^2)+ 
\frac{\left( 1+T^V(-\mu^2)G_0(-\mu^2)\right)|\chi\rangle
\langle\chi|\left(1+G_0(-\mu^2)T^V(-\mu^2)\right)}
{\displaystyle \frac{2\pi^2}{\lambda_{\delta}}-
\langle\chi|G_0(-\mu^2)|\chi\rangle -
\langle\chi|G_0(-\mu^2)T^V(-\mu^2)G_0(-\mu^2)|\chi\rangle} 
\ , \label{19a}
\end{eqnarray}
}where the denominator at an arbitrary energy $-\mu^2$, can be defined 
by a constant $C^{-1}(-\mu^2)$, 
{\small\begin{equation}
2\pi^2C^{-1}(-\mu^2)\equiv
\frac{2\pi^2}{\lambda_{\delta}}-\langle\chi
|G_0(-\mu^2)\left[1+T^V(-\mu^2)G_0(-\mu^2)\right]|
\chi\rangle
\equiv\frac{2\pi^2}{\lambda_{\delta}}-\langle\chi|G_V(-\mu^2)|\chi\rangle ,\label{16b}
\end{equation}
}where $G_V(-\mu^2)=-\left(\mu^2+H_0+V\right)^{-1}$ is the free propagator associated 
to the regular potential. 
The fixed-point structure of the renormalized potential (\ref{15})
defines the functional form of $C(-\mu^2)$ at the subtraction point.
By moving the scale to $\mu^{\prime}$, as $\partial V_{\cal R}/\partial\mu^2=0$,
we have
\begin{equation} 
2\pi^2C^{-1}(-\mu^{\prime 
2})-2\pi^2C^{-1}(-\mu^{2})=\langle\chi|\left(
G_{V}(-\mu^{2}) - G_{V}(-\mu^{\prime 2}) 
\right)|\chi\rangle .
\end{equation} 
By choosing $\mu^2$ at one of the binding
energies of the physical system, $\mu^2=\mu^2_B$ and  
$C^{-1}(-\mu_B^2)=0$. 
So, from Eq.~(\ref{19a}), with the cutoff $\Lambda$ included 
via step function $\Theta(x)$ ($=0$ for $x<0$ and $=1$ for $x>0$), 
{\small\begin{equation}
\frac{1}{\lambda_{\delta}}=
-\frac{1}{2\pi^2}\int d^3p
\frac{\Theta(\Lambda^2-p^2)}{\mu_B^2+p^2}\left[1
 - \int d^3q
\frac{\Theta(\Lambda^2-q^2)}{\mu_B^2+q^2}
{\langle \vec{p} |T^V(-\mu_B^2)| \vec{q} \rangle}\right] .
\label{16}
\end{equation}
}With $\Lambda$ at the exact infinite limit,
$\lambda_{\delta}$ contains the divergences in the momentum integrals, 
canceling the infinities of Eq.~(\ref{2}). It should be clear that, 
the role of the cutoff parameter $\Lambda$ is just to provide a regulator 
for the integrals. At the end, it should disappear in the exact $\infty$ 
limit, without affecting the physical results.
The relevant scale parameter, where the physical information are supplied, is
the energy-point $-\mu^2_B$. As specific choice of 
the subtraction point position should not affect the results.

Given a specific example for two identical particles,
with $V_{\cal R}$ being the interaction in the matrix
elements given in Eq.~(\ref{15}), we can follow 
by the numerical diagonalization of the fixed-point Hamiltonian, Eq.~(\ref{1}), 
to obtain the associated bound-states $\varepsilon$. With the 
system in the $s-$wave, in units such that $\hbar^2/(2m)=1$ ($m$ the mass 
the particles), 
the corresponding Schr\"odinger equation can be written as 
{\small\begin{eqnarray}
H_{\cal R}\Psi(p)= p^2\Psi(p)+ 
{\frac{2}{\pi}}\int_0^{\Lambda} q^2 dq 
\left[\frac{-1}{2pq}\ln\left(\frac
{\eta^2+(p+q)^2}{\eta^2+(p-q)^2}
\right)+\lambda_\delta\right]
\Psi(q) 
=\varepsilon\Psi(p) \ ,
\label{20}
\end{eqnarray}
}where $\lambda_\delta$ is provided by the 
renormalization prescription, keeping $\mu_B^2$ fixed by  
one of the bound-states.

In order to become more clear the approach, let us first consider an exact 
numerically soluble system, described by an arbitrary {\it reference potential}, 
with the corresponding matrix elements given by
\begin{equation}
\langle\vec{p}|V|\vec{q}\rangle= -\frac{1}{\pi^2} \left[
\frac{1}{|\vec{p}-\vec{q}|^2 + \eta^2}+ 
\frac{1}{|\vec{p}-\vec{q}|^2 + \eta_S^2}
\right],\label{refpot}\end{equation}
where we assume $\eta^2=0.01$ and $\eta_S^2=1$, considering 
all energy dimensional quantities ($\eta^2$, $\eta^2_S$, as well as
$\varepsilon$) in units of inverse-squared length. 
This reference potential produces three bound-state energies: 
$\varepsilon^{(0)}=-$2.3822, $\varepsilon^{(1)}=-$0.20297 and 
$\varepsilon^{(2)}=-$0.020643. Next, to verify how the renormalization approach
works when the interaction has a singular term, but still describing the same 
physics, let us assume that the short-range part of the reference potential is 
replaced by a Dirac-delta function  with the strength $\lambda_\delta$, such that 
$\langle\vec{p}|V|\vec{q}\rangle\to\langle\vec{p}|V_{\cal R} |\vec{q}\rangle$.
This total interaction, with regular long-range and singular short-range
parts, is renormalized by assuming a physical constraint supplied by one of 
the bound-state energies, which is supposed to be known in our hypothetical 
example. Therefore, the subtraction point $-\mu^2$, provided by this energy, 
is regularizing the formalism, via a subtraction procedure, as well as
carrying the relevant physical information in the present Hamiltonian 
renormalization approach.

\begin{figure}
\begin{center}
\hspace{-1mm}
\includegraphics[scale=0.4]{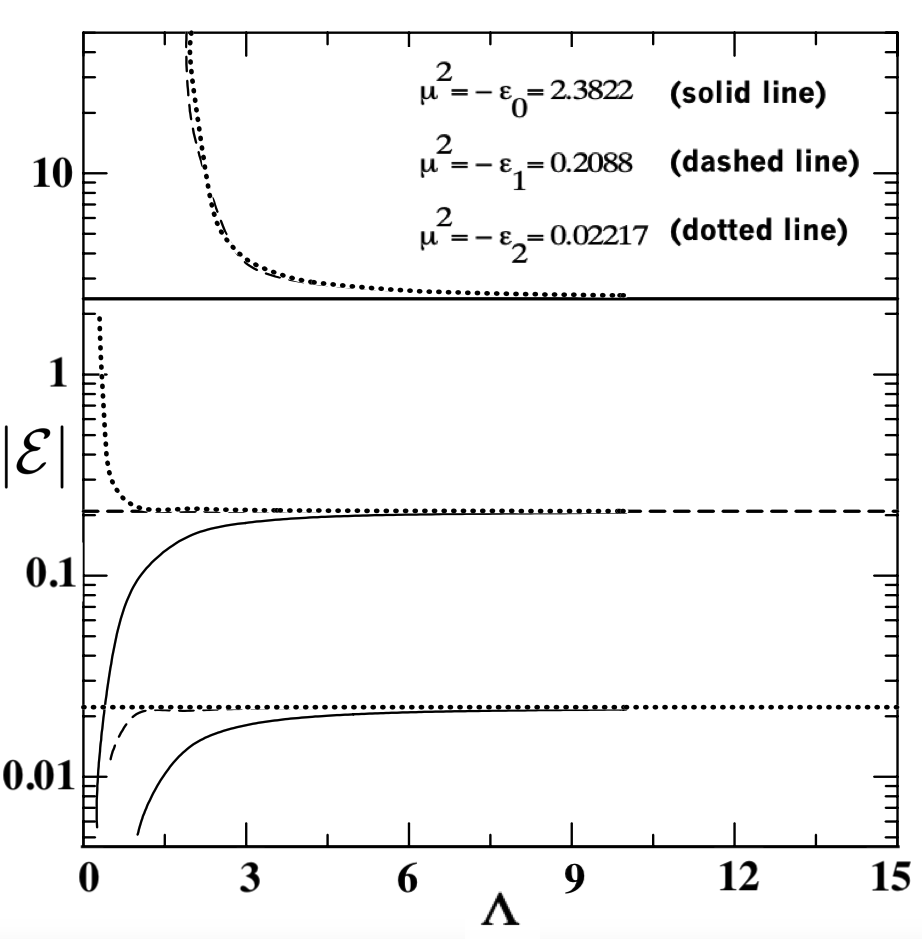}
\end{center}
\caption{
The convergence of the numerical results for the bound-state energies
$\varepsilon$ is verified for the Hamiltonian given in Eq.~(\ref{20}),
by considering our described renormalization procedure. The three exact 
energy eigenvalues are shown by the horizontal lines, as indicated.
The same exact results are obtained, in the momentum cut-off 
limit $\Lambda\to\infty$, irrespectively to the values of the energy-scale
parameter $-\mu^2$, which are used to renormalize the theory.
The solid curves (converging to the first- and 
second-excited states) are given by assuming $\mu^2=2.3822$.
The dashed curves (converging to ground and second-excited states) are
for $\mu^2=0.2088$. The dotted curves (converging to ground and 
first-excited states) are for $\mu^2=0.02217$. All energy values
($\mu^2$, $\Lambda^2$ and $\varepsilon$) are in inverse-squared length units.}
\label{fig:1}       
\end{figure}

In Fig.~\ref{fig:1}, we are presenting the corresponding numerical results 
obtained for the eigenvalues of the Hamiltonian (\ref{20}), as functions of 
the momentum cutoff parameter $\Lambda$. The three eigenvalue energies obtained
by considering the reference regular potential (\ref{refpot}) are shown
by the three horizontal lines.
As verified, the specific choice of the subtraction point $-\mu^2$ 
(one of the three values shown by the straight lines) does not affect 
the final convergent results, which are exact in the limit $\Lambda\to\infty$. 
The Fig.~\ref{fig:1} displays three sets of results, such that, 
for each one, the value of $-\mu^2$ is specifically defined by
one of the assumed known energies.
Wth $\mu^2=$ $-\varepsilon^{(0)}=$ 2.3822, the results are given with
solid lines for the first and second excited energies. In this
case, the exact results obtained from Eq.~(\ref{15}) with 
$\Lambda\to\infty$ are $\varepsilon^{(1)}=-$0.2088 and 
$\varepsilon^{(2)}=-$0.02217. 
The results for the other two sets are obtained by using the
same procedure: By assuming $\mu^2=$ $-\varepsilon^{(1)}=$ 
0.2088, the results are with dashed lines;  and when 
$\mu^2=$ $-\varepsilon^{(2)}=$ 0.02217, they are represented by
dotted-lines. As verified, this diagonalization procedure provides 
stable results when $\Lambda\to\infty$, converging to the exact 
values, given by the real poles of the T-matrix: 
$\varepsilon^{(0)}=-$2.3822, $\varepsilon^{(1)}=-$0.2088 and 
$\varepsilon^{(2)}=-$0.02217.

As the renormalized Hamiltonian does not depend on the choice of $\mu$,
it is a fixed-point Hamiltonian in this respect.
The above example is providing a clear picture about what we have 
stated in Eq.~(\ref{11}). The 
momentum cutoff $\Lambda$ is just an instrumental regulator, which 
disappears  as a natural infinite limit of the integrals, 
where all the infinities presented in the formalism are canceled out.

\subsection{Four-term-singular renormalized Hamiltonian}\label{subsec3.2}

A four-term-singular bare interaction is considered here in order to derive the 
explicit form of the renormalized potential, obtained for the $s-$wave
after partial-wave decomposition. The matrix 
elements of this bare singular potential, in terms of powers 
of the momentum, is given by
\begin{equation}
\langle p|V|q\rangle=\sum_{i,j=0}^{1}\lambda_{ij} p^{2i} q^{2j}
\;\;\;(\lambda_{ij}=\lambda_{ji}^*) \ ,  \label{22}
\end{equation}
with the renormalized strengths fixed by
the physical scattering amplitude at a reference energy $-\overline\mu^2$, 
\begin{eqnarray}
\langle p| T(-{\overline\mu}^2)|q\rangle=\lambda_{{\cal R}00}
 +\lambda_{{\cal R}10} (p^2+q^2)+\lambda_{{\cal R}11} p^2 q^2 \ .  
\label{23}
\end{eqnarray}
For simplicity, we assume that all strengths $\lambda_{{\cal R}ij}$ are real 
to have a Hermitian renormalized Hamiltonian. 
With the bare interaction given in~(\ref{22}), the physics of the 
system becomes completely defined by the values of the 
renormalized strengths, $\lambda_{{\cal R}ij}$, obtained
at the reference energy $-\overline\mu^2$, which is also part of the physical input.
Here, the units are $\hbar=m=1$, with $m$ the particle mass.

The potential given by Eq.~(\ref{22}) implies in integrals that diverge 
at most as $p^5$.
In order to obtain finite integrals, this requires at least 
three subtractions in the kernel of the corresponding 
LS equation. With $n=3$ in Eq.~(\ref{7}), from the 
recurrence relationship (\ref{4}), the following equations
are derived:
\begin{eqnarray}
&&\langle p|V^{(1)}(-\overline{\mu}^2)|q\rangle =\lambda_{{\cal R}00},
\quad \langle p|V^{(2)}(-\overline{\mu}^2;k^2)|q\rangle =
\frac{1}{\lambda_{{\cal R}00}^{-1}+ I_0},\label{24}\\
&&\langle p|V^{(3)}(-\overline{\mu}^2;k^2)|q\rangle = 
\overline{\lambda}_{{\cal R}00} 
+\lambda_{{\cal R}10}(p^2+q^2) 
+\lambda_{{\cal R}11} p^2q^2,\nonumber
\end{eqnarray}
where ${\overline\lambda}_{{\cal R}00}\equiv
\left[{\lambda_{{\cal R}00}^{-1}+ I_0+I_1}\right]^{-1},$ with 
$I_{i=0,1}\equiv I_i(k^2,\overline{\mu}^2)$ defined by
\begin{eqnarray}
&&I_i(k^2,\overline{\mu}^2) \equiv
\frac{2}{\pi}\int^\infty_0dqq^2
\frac{(\overline{\mu}^2+k^2)^{1+i}}{(\overline{\mu}^2+q^2)^{2+i}} =
\frac{(\overline{\mu}^2+k^2)^{1+i}} {(2\overline{\mu})^{1+2i}}\ 
\label{25}.\end{eqnarray}
Note in above that the singular terms, as shown in Eq.~(\ref{4}),
are introduced for $n=3$ in $V^{(3)}$.
Also, we noticed that $I_i=0$ when $k^2=-\overline\mu^2$.
By introducing $V^{(3)}(-\overline{\mu }^{2},k^{2})$ 
of Eq.~(\ref{24}) in Eq.~(\ref{7}), the renormalized interactions 
are obtained analytically, in this example, with the strenghts 
$\Lambda _{ji}(k^2)$ not depending on the subtraction point, given by:

\begin{eqnarray}
\langle p|V_{{\cal R}}|q\rangle =\sum_{i,j=0}^{1}\Lambda
_{ij}(k^{2})p^{2i}q^{2j}\;\quad  {\rm where}\quad
[\Lambda _{ij}(k^2)=\Lambda _{ji}(k^2)]\quad {\rm with}
\label{26}
\end{eqnarray}
{\small\begin{eqnarray}
\Lambda _{00}(k^{2})&=&
\frac
{\overline{\lambda }_{{\cal R}00} -
\left(\overline{\lambda}_{{\cal R}00} K_{1} +
      \lambda _{{\cal R}10}K_{2}\right)\Lambda _{10}(k^{2})}
{1+\overline{\lambda}_{{\cal R}00}K_{0}+\lambda_{{\cal R}10}K_1
} ;\quad  
\Lambda _{11}(k^{2})=
\frac
{\lambda_{{\cal R}11} -
\left(\lambda_{{\cal R}10} K_{0} +
      \lambda _{{\cal R}11}K_{1}\right)\Lambda _{10}(k^{2})}
{1+\lambda_{{\cal R}11}K_2+\lambda_{{\cal R}10}K_1 
};\nonumber  \\
\Lambda_{10}(k^{2})&=&
\frac{\lambda_{{\cal R}10}+
(\lambda_{{\cal R}10}^{2}-\overline{\lambda }_{{\cal R}00}
\lambda _{{\cal R}11})K_{1}} 
{1+ \overline{\lambda}_{{\cal R}00}K_{0}+\lambda_{{\cal R}10}K_1 
+\lambda_{{\cal R}11}K_2+\lambda_{{\cal R}10}K_1
+(\lambda_{{\cal R}10}^2 - \overline{\lambda }_{{\cal R}00}
\lambda_{{\cal R}11})(K_1^2-K_0 K_2) 
},  \;\;\;\label{29}
\end{eqnarray}
}where 
\begin{eqnarray}
K_{i=0,1,2}&\equiv&K_{i}(k^{2},{\overline{\mu }}^{2})
\equiv \frac{2}{\pi}
\int_{0}^{\infty }\frac{dqq^{2i+2}}{k^2-q^2}
\left[1- \left(\frac{\overline{\mu}^2+k^2}
{\overline{\mu}^2+q^2}\right)^3\right]
.\label{31} \end{eqnarray}
The $K_{i}$ are the divergent integrals, which cancel exactly the
infinities of the LS equation obtained with the 
renormalized interaction (\ref{26}).
The integrands are given by the kernel of
Eq.~(\ref{7}) with $n=3$.

In view of the arbitrariness of the subtraction point, the values of 
$\Lambda_{ij}(k^2)$ are independent on the scale $\mu$, with 
${\partial \Lambda_{ij}(k^2)}/{\partial\mu^2}=0.$
These conditions on the derivatives are given by the explicit form of 
(\ref{11}) in the case of the four-term-singular potential. However,
the evolution of the driving term $V^{(3)}(-\mu^ 2;k^ 2)$ 
with $\mu$ can be computed by solving the first order differential CS 
equation~(\ref{13}) with the boundary condition at the initial scale 
$\mu^2=\overline \mu^2$. This would 
imply in a nontrivial dependence of  the coefficients 
$\overline\lambda_{{\cal R}00}$, 
$\lambda_{{\cal R}10}$ and $\lambda_{{\cal R}11}$ with $\mu^2$ and $k^ 2$, 
which after all, will keep unchanged the $T-$matrix from the solution of the 
third-order subtracted scattering equation~(\ref{3}) using the new subtraction point.

\subsection{Subtracted renormalization scheme for the 
one-pion-exchange potential}

As a third pedagogical example, we consider here the application of 
the renormalization scheme, based on the subtracted T-matrix equation, 
to the neutron-proton system, with the basic formalism recovered from
Ref.~\cite{1999fred}, which was based in Weinberg's pioneering 
work~\cite{1990wein}. 
The subtraction parameter $\mu$ can run to infinite, which  was indeed 
verified in the $^3$S$_1-\,^3$D$_1$ and $^3$S$_0$ neutron-proton channels 
with the one-pion-exchange potential (OPEP) supplemented by contact 
interactions.  

For the unregulated effective potential, in the corresponding matrix elements, 
a power expansion in the mid- and short-range parts of the interaction is 
usually assumed, in order to keep intact the well established long-range part 
of the OPEP.
So, the matrix elements of the full effective nucleon-nucleon (ENN) interaction, 
with ${\vec p}- \vec q$ being the momentum transfer,
can be expressed by 
\begin{eqnarray}
\langle{\bf p}|V_{ENN}|\vec q \rangle
= \langle\vec {p}|V_\pi|\vec q \rangle &+& \frac{1}{2\pi^2}
\left[
\frac{1 - \vec\tau_1\cdot\vec\tau_2}{4}
\left( \lambda_t^{(0)} + \lambda_t^{(1)}q^2
 + \lambda_t^{(1)*}{p}^2 + \cdot \cdot \cdot \right) 
+ \right. \nonumber \\
&+& \left. \frac{3+\vec\tau_1\cdot\vec\tau_2}{4}
\left(\lambda_s^{(0)} +\lambda_s^{(1)}q^2 
+ \lambda_s^{(1)*}{p}^2 + \cdot \cdot \cdot \right) 
\right], \label{veft}
\end{eqnarray} 
with
\begin{eqnarray}
\langle\vec {p}|V_\pi|\vec q \rangle
=-\frac{g_a^2}{4(2\pi)^3f_\pi^2} \vec\tau_1\cdot\vec\tau_2 
\frac{\vec\sigma_1\cdot(\vec {p}-\vec {q})\;  
 \vec\sigma_2\cdot(\vec {p}-\vec {q})}
{ (\vec{p}-\vec{q})^2+m_\pi^2}\ ,
\label{opep}
\end{eqnarray} 
where $\sigma_i$ and $\tau_i$ are the usual spin and isosping Pauli
matrices for the nucleon $i=1,2$. The subindices $t$ and $s$ in the
expansion strength parameters are for spin triplet (isospin singlet) 
and spin singlet (isospin triplet) channels, respectively. 
$g_a\;(=1.25)$ is the axial coupling constant,
with $f_\pi \; (=93$ MeV) the pion weak-decay constant, with 
$m_\pi\; (=138$ MeV) the pion mass. 
By assuming only the leading order (LO) term of the expansion~(\ref{veft}), 
we have $\lambda_{s(t)}^{(n)} = 0$ for all $n\ge 1$. 

\begin{figure}
\begin{center}
\hspace{-1mm}
\includegraphics[scale=0.365]{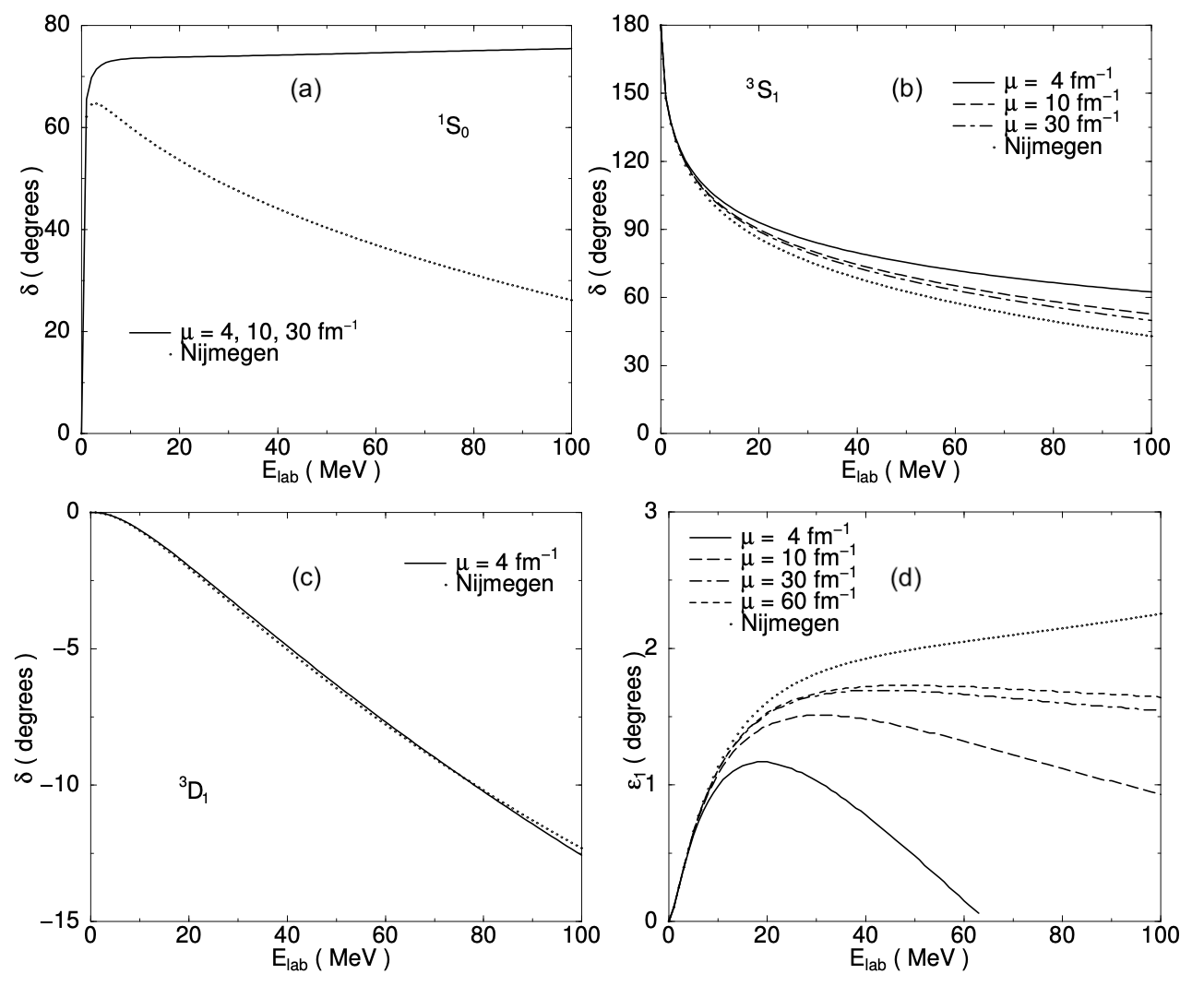}
\end{center}
\caption{
Neutron-Proton phase shifts for the $^1S_0$ state, $\delta_{0,s}$ 
[panel (a)]; $^3S_1$ state, $\delta_{0,t}$ [panel (b)]; $^3D_1$ state  
$\delta_{2,t}$ [panel (c)]; and mixing parameter for the $^3S_1-^3D_1$ 
states[panel (d)], according to Ref.~\cite{stapp}. 
The model results are shown for different values of  $\mu$, 
as identified inside 
the panels. The dotted lines are data results  
from Ref.~\cite{nij}. These four-panel are extracted from plotted 
results presented in Ref.~\cite{1999fred}. With more than one
subtraction, see Ref.~\cite{2011tim}.}
\label{fig:2}       
\end{figure}

Basically, it was applied in this case the same procedure as presented
before, for the subtracted T-matrix formalism and for the 
renormalization of the interactions, with details also given in  
Ref.~\cite{1999fred}. For that, only one scaling parameter $\mu$ was
used, with the corresponding results showing an overall agreement 
with the neutron-proton data, particularly for the observables related to the 
triplet channel at low energies. The agreement is qualitative in the $^1$S$_0$ channel. 
These results are shown in the four panels of Fig.~\ref{fig:2}. The mixing
parameter for the $^3S_1-^3D_1$ states, considering the definition given
in Ref.~\cite{stapp}, was verified to be the most sensible observable 
to the scale [see panel (d) of Fig.~\ref{fig:2}].
However,  we should observe that the renormalization procedure, with just one subtraction and the triplet scattering length kept fixed, is enough to show a converged mixing parameter in the limit of $\mu\to \infty$,  as panel (d) of Fig.~\ref{fig:2} indicates.

With the renormalization group invariant approach for the subtraction 
procedure providing the basis on how to derive fixed-point Hamiltonians,
this example was followed by a few other applications related to the OPEP, 
such as the corresponding procedure for the next-to-leading order (NLO) 
nucleon-nucleon interaction~\cite{2007tim}. Within a more detailed methodology to 
renormalize the two nucleon interaction, with no need of cutoff regularization, 
by including more than one subtraction, better fits are provided for 
the nucleon-nucleon scattering observables in Ref.~\cite{2011tim}.

In particular,  we conclude this one-pion-exchange potential example, by observing that the 
applicability  of the subtractive renormalization procedure can be further
extended to cases in which higher order singularities exist in  the 
interactions.  By following the arguments of a similar approach developed in 
Ref.~\cite{2008bira},  applied to renormalization of singular potentials containing divergences $1/r^2$ and $1/r^4$, 
one  verifies that the present approach is perfectly suitable to the case in which we have the 
tensor part in spin-triplet channels of the one-pion-exchange interaction, which goes with  
$r^{-3}$ at the origin~\cite{2005Nogga}, as the example shown in Fig.~\ref{fig:2}, where the OPEP 
in Eq.~(\ref{opep}) includes such singularity. As also pointed out in Ref.~\cite{2011tim}, the number of recursive steps 
required to renormalize the interaction depends on how the potential diverges, such that 
the method developed in Ref.~\cite{2000frederico} can be implemented with a generic number of 
recursive steps or subtractions.

\section{Three-body subtracted equations with renormalized 
Hamiltonian}\label{sec4}
 In this section, we review the derivation of the subtracted three-body 
 Faddeev $T-$matrix equations proposed in~\cite{1995Adhikari1} 
 and the renormalized Hamitonian, which contains two and 
 three-body potentials. For a more detailed discussion, related to  
 the subtracted formalism applied to three-body neutron-halo structures, 
 which includes previous contributions, see Ref.~\cite{2012fred}. 
A microscopic study
of the trimer scaling function, within the perspective to be verified
in cold-atom experiments, was done more recently in Ref.~\cite{2021madeira}.
Within the effective field theory (EFT),  halo nuclei systems were also 
reviewed in~\cite{HammerJPG2017}.
In the scattering region, the neutron-deuteron scattering problem has 
been solved with a different form of subtracted equations~\cite{AfnPRC2004}.
The neutron-$^{19}$C
$s-$wave elastic scattering was studied in~\cite{YamPLB2008b,ShalchiPLB2017}.
In atomic physics, the subtracted scattering equations were applied 
recently to different scattering problems, like the Efimov discrete 
scaling~\cite{Efimov,EfimovNPA1981} in an atom-molecule collision~\cite{ShalchiPRA2018}  
and to study cold atom-dimer reaction rates with $^4$He, $^{6,7}$Li, 
and $^{23}$Na~\cite{ShalchiPRA2020}. The application of subtracted equations 
to the four-boson problem with Dirac-delta interaction was proposed in 
Ref.~\cite{YamEPL2006}, being further explored some years later leading to 
the discovery of a new four-boson limit cycle~\cite{HadPRL2011},
independent of the Efimov discrete scaling.

\subsection{Subtracted Faddeev Equations}

The subtraction method used to define the two-body T-matrix reviewed 
in section~\ref{sec2} was generalized to three-body systems. The 
subtracted Faddeev equations for the T-matrix~\cite{1995Adhikari1} 
were introduced in the context of a zero-range potential, although 
it can be generalized to account 
for potentials with a short-range term plus a Dirac-delta. The  
subtracted Faddeev equations follows the method applied to the two-body 
LS equation, where now the three-body free Green's function in the  
kernel of the Faddeev integral equations are regularized by the 
subtraction $G_0(E)-G_0(-\mu_3^{2})$ substituting the free three-body 
Green's function $G_0(E)$. As detailed in the case of the two-body 
problem, the advantage of using the subtracted T-matrix equations 
relies on its explicit renormalization group invariance, and the 
possibility of defining an associated renormalized Hamiltonian.

The three-body $T-$matrix at the subtraction  point $-\mu_3^2$ is 
the sum of the two-body $T-$matrices for all the 
subsystems derived for singular potentials, which is given by
\begin{eqnarray}
T(-\mu_{3}^2)=\sum_{k=1,2,3} t_{(ij)}
\left(-\mu_{3}^{2}-\frac{q_{k}^2}{2\;m_{ij,k}}\right), 
\label{twobt}
\end{eqnarray}
where 
$k=1,2,3$ refers to one of the particles, with the remaining
interacting pair $ij$ corresponding cyclically to $k$, as 
$ij=23,31,12$. For a given $k$, 
$m_{ij,k}$ is the associated reduced mass between this particle 
and the mass of the $ij-$pair $m_i+m_j$, given by 
$m_{ij,k}\equiv m_k(m_i+m_j)/(m_i+m_j+m_k)$. The two-body 
$T-$matrix $(t_{(ij)})$ is evaluated at the subsystem energy in 
the three-body center-of-mass. 

 The set of subtracted Faddeev equations, which have their 
 detailed derivation in~\cite{2012fred},
 are given by:
\begin{equation}
T_k(E)=t_{(ij)}\left(E-
\frac{q_{k}^2}{2m_{ij,k}}\right)\left[1+\left(G_0^{(+)}(E)-
G_0(-\mu_{3}^2)\right)\left(T_{i}(E)+T_{j}(E)\right)\right].
\label{tfaddeev}
\end{equation}
The solution of the homogeneous form of the above set of equations 
gives the bound-state energy, with the associated Faddeev component of 
the wave function vertex. In the case of the Dirac-delta potential 
the coupled set of equations is the regularized form~\cite{2012fred} of the 
zero-range Skorniakov and Ter-Martirosian (SKTM) equation for the three-boson bound 
state~\cite{Skorniakov1957}. When $\mu_3\to\infty$, we have the occurrence 
of the Thomas collapse~\cite{ThomasPR1935}.  However, for finite $\mu_3$ the
scale invariance of the zero-range three-body T-matrix equation in
the ultraviolet momentum region is broken and the Thomas collapse 
in the $s-$wave state of maximum symmetry does not happen. The 
correlations between three-body observables in this state tend to 
achieve a limit cycle,  where the dependence on $\mu_3\to \infty$ 
does not matter and the theory in this sense is fully renormalized.

\subsection{Renormalized Three-body Hamiltonian}

The renormalized three-body interaction Hamiltonian~\cite{2012fred}, 
$H^{(3)}_{{\cal R}I}$, is obtained from Eq.~(\ref{7}). By 
resorting to one subtraction ($n=1$) where, at the subtraction 
point $-\mu_3^2$, we can identify $V^{(1)}$ with Eq.~(\ref{twobt}). From 
that, we can write the following equivalent equation:
\begin{eqnarray}
H^{(3B)}_{{\cal R}I}=\sum_{k}\left[
t_{(ij)}\left(-\mu_{3}^2-\frac{q^2_{k}}{2m_{ij,k}}\right)\right]
\left(1-G_0(-\mu^2_{3})H^{(3B)}_{{\cal R}I}\right) \ .\label{v3ren1}
\end{eqnarray}
The renormalized three-body interaction Hamiltonian~(\ref{v3ren1}) 
can be split in two- and three-body ones, by introducting the
Faddeev decomposition of the three-body potential was introduced, as
\begin{equation}
H^{(3B)}_{{\cal R}I}
=\sum_{k}\left[ V^{(2B)}_{{\cal R}(ij)}+V^{(3B)}_{{\cal R}(k)}\right]
\ , \label{h3i}
\end{equation}
where, after formal manipulations, one gets the solution, which is given 
in matricial format as
\begin{eqnarray}
\left[
\begin{array}{ccc}
1 & V^{(2)}_{{\cal R}(23)}G_0(-\mu_{3}^2) & V^{(2)}_{{\cal R}(23)}G_0(-\mu_{3}^2) \\
V^{(2)}_{{\cal R}(31)}G_0(-\mu_{3}^2)& 1 & V^{(2)}_{{\cal R}(31)}G_0(-\mu_{3}^2) \\
V^{(2)}_{{\cal R}(12)}G_0(-\mu_{3}^2)&V^{(2)}_{{\cal R}(12)}G_0(-\mu_{3}^2)& 1
\end{array}
\right] 
\left[
\begin{array}{c}
V^{(3)}_{{\cal R}(1)}+ \\
V^{(3)}_{{\cal R}(2)}+\\
V^{(3)}_{{\cal R}(3)}+
\end{array}
\begin{array}{c}
V^{(2)}_{{\cal R}(23)} \\
V^{(2)}_{{\cal R}(31)}\\
V^{(2)}_{{\cal R}(12)}
\end{array}
\right]=
\left[
\begin{array}{c}
V^{(2)}_{{\cal R}(23)} \\
V^{(2)}_{{\cal R}(31)}\\
V^{(2)}_{{\cal R}(12)}
\end{array}
\right]
\cdot\label{kernel3}
\end{eqnarray}
Note that  $V^{(3B)}_{{\cal R}}$ is fully connected and contains 
the boundary condition at the subtraction point, Eq.~(\ref{twobt}), necessary for 
building the subtracted $T-$matrix equation for the three-body system.
In the case of the Dirac-delta interaction, where one needs to solve the  
SKTM equations, the three-body renormalized 
potential allows its solution by introducing a subtraction in the kernel. 
This is the counterpart  of the three-body potential necessary in the 
EFT approach~(see,  e.g., Ref.~\cite{2020ham}) to solve the SKTM equations. 
Therefore, the limit cycles and the scaling functions, which express 
correlations between three-body observables, obtained from the
subtraction method, should agree with the corresponding approach
derived from EFT applied to the three-body problem.

\section{Four-body scale and subtracted equations}\label{sec5}

The subtraction method used to define the two- an three-body T-matrix was 
also introduced in the Faddeev-Yakubovski (FY) formalism when considering 
four-particle systems with the Dirac-delta interactions in 
Ref.~\cite{YamEPL2006}, which was followed by more detailed investigations
by some of us 
in Refs.~\cite{YamEPL2006,HadPRL2011,FredFBS2011,HadPRA2012,FredFBS2012}. 
In the four-bosons case, besides the  three-body subtraction scaling point, 
another subtraction point has to be introduced, directly associated with 
the four-body scale, due to new terms in the FY coupled integral equations, 
which are not directly identified with the three-body kernel. 

Next, we describe the main ideas concerned the application of the 
renormalization subtraction method, which was proposed in Ref.~\cite{YamEPL2006}, 
to the four-body case. By considering the four particles identified by $i,j,k,l$
ranging from 1 to 4, the FY components of the wave function associated with the 
3+1 partitions [$(ijk)+l, (jki)+l$ and $(kij)+l$], among the 18 possibilities, 
will fully describe the three-body subsystems $(ijk)$, such that,  when the 
interaction with the fourth particle is turned off, the FY equations should 
reduce to the usual Faddeev three-body ones. With this reasoning, the subtracted 
form of the free Green's function at the subtraction point $-\mu_3$ is introduced, 
as shown for the Faddeev equations (\ref{tfaddeev}). Physically, for the 
three-boson case, this subtraction point is associated with the necessity 
of an independent three-body scale to define the observables in the zero-range 
interaction limit, leading to limit cycles for the correlations between two 
observables of the three-boson S-wave state.
Therefore, in the free Green's function, which appears together with 
the FY components, associated to the  three-body subsystem, it 
is adopted the energy subtraction at $-\mu_{3}^{2}$:
\begin{equation}
G_0(E) \longrightarrow G_{0}^{(3)}(E) \equiv
G_0(E)-G_0(-\mu_{3}^{2}). \label{G03}
\end{equation}
In principle, a different energy subtraction parameter $-\mu_4^2$, should emerge in 
the subtracted form of the Green's functions coming together with  the remaining 
fifteen FY components of the wave function. By following this reasoning, it was 
introduced in~\cite{YamEPL2006} the subtraction of those Green's function as:
\begin{equation}
G_0(E) \longrightarrow G_{0}^{(4)}(E)\equiv G_0(E)-G_0(-\mu_{4}^{2})\, .
\label{G04}\end{equation}
This new subtraction would be irrelevant if one let $\mu_{4}\to\infty$ without 
consequences. However, it was observed in \cite{YamEPL2006} that the four-boson 
ground state collapses in this limit. Only in Ref.~\cite{HadPRL2011} it was 
recognized that a new four-boson limit cycle occurs, which is being interwoven 
with the three-body Efimov limit cycle~\cite{TomioFBS2014}. This limit cycle is  
associated with a new discrete scaling factor. It appears in the scaling 
function associated with the correlation between two consecutive tetramer 
energies at the unitary limit for a fixed trimer energy. The so-far results 
described here were also found consistent with the ones obtained in 
Ref.~\cite{DeltuvaFBS2011}, which were obtained by considering finite-range 
potentials, corroborating the above analysis related to four-boson systems.  
By addressing general aspects of the universality in few-body systems, which
include some discussion beyond three-body systems, we have already a few 
reviews which have appeared in the last decade, such 
as Refs.~\cite{ZinnerJPG2013,NaidonRPP2017,GreeneRMP2017}.

The new four-boson limit-cycle has a discrete scaling which differs from the 
three-boson Efimov factor, corresponding to the breaking of the continuous 
scale symmetry to a discrete one verified in the FY zero-range 
equations~\cite{FredFBS2019}. This brings together the necessity of a new 
four-boson scaling factor. The analytical derivation detailed 
in Ref.~\cite{FredFBS2019} provides a discrete ratio different from the Efimov 
one, given by $s_4$, with the corresponding transcendental equation being
such that the discrete ratio between the energies of successive tetramer 
states is given by $\rm{e}^{-2\pi/s_4}$, in the $\mu_4\to\infty$ limit for 
fixed trimer energy. 
This analysis, in which a four-boson scale emerges, being associated to a 
limit cycle, was further supported by the recent study in 
Ref.~\cite{dePaulaJPG2020}, in which a system with $N$-light bosons and 
two heavy ones are considered within the Born-Oppenheimer approach. In
this case, it was found that the strength of the attractive $1/r^2$ 
interaction depends on the number $N$ of light bosons, in correspondence 
to $s_4$. The study was done for the particular case in which the 
interactions (contact ones) can occur only between the light particles 
with the heavy ones. 
Therefore, in the case of the four-boson system, our expectation is that 
a four-body potential should emerge associated with the evolution of the 
system properties due to the new scale. However, providing an emergent 
universal behavior independent of the Efimov one. 

Recently,  it was demonstrated the necessity of the four-boson scale  
within the context of the EFT at  NLO  in Ref.~\cite{BazakPRL2019}.
Such finding should be reconciled  with the  
breaking of the continuous scale symmetry of the FY equations 
to a discrete one in the limit of a zero-range potential, as expressed by the 
correlation found for the tetramer energies~\cite{HadPRL2011}. The evolution 
along the correlation plot is implicitly governed by a four-body interaction, 
which would be in principle related to the subtraction in the FY equation, 
in this sense tuning the EFT four-body potential at NLO one eventually could 
find the trace of such correlation. 

To close this section, we mention that other approaches have studied the 
universality and scaling in $N$-boson systems, with short-range interactions, 
as in Refs.~\cite{GattobigioPRA2014,KievskyPRA2014}, where it was also  
considered the $N$-boson spectrum~\cite{KievskyPRA2014}, 
without short-range four-boson forces, in which it was not possible 
to identify the dependence on the scales beyond the three-body one. 

\section{Conclusions}\label{sec6}
In summary, in this contribution to the memory of Steven Weinberg,
we are reporting some works we have developed, which were mainly inspired 
in the fundamental contributions of Weinberg to the few-body physics, 
considering effective interactions among two- and three-nucleon systems. 
We start the report by considering the general ideas and works related to 
effective interactions which contains short-range singularities, from which
concepts as universality and limit-cycles follow from the renormalization 
group approach. In section 2, we provide the basic formalism related to 
effective interactions, described by a Hamiltonian renormalization 
approach, which emerges as a consequence of
a renormalization procedure, applied to the corresponding scattering matrix,
at a fixed-point energy scale. 
The approach, shown to be renormalization group invariant, 
is based on a subtraction procedure, from which a fixed-point Hamiltonian
can be derived. This renormalized Hamiltonian is taken as a fixed-point
operator, in the sense that it does not depend on the position of the 
subtraction point $-\mu^2$, where the physical information is supplied 
to the theory.
It naturally includes the renormalization group invariance properties 
of quantum mechanics with singular interactions, as expressed by the 
non-relativistic Callan-Symanzik equation.  
The theory is supplemented by three examples of applications to 
the case of two-particle systems, which have one or more singularities 
in  their original interactions.

In section 4, we show how to apply the subtracted renormalization 
approach to three-body systems through the Faddeev 
formalism with singular two-body interactions. The section is
concluded with some details on the possible extension to 
systems with four or more particles.
The wide range of applicability of renormalized Hamiltonians, from atomic 
to nuclear physics models derived from effective theories of the QCD, is
also emphasized in this section.

As a perspective, it would be of interest a comparison between the 
non-perturbative Hamiltonian renormalization approach, described in this report,
with other available renormalization techniques applied to quantum few-body 
systems; such as, for example, the approach considered in Ref.~\cite{birse}. 
However, a caution is necessary when doing such comparison, as one should note 
that, in the present work, the invariance of the Hamiltonian is with respect 
to a subtraction energy scale, in the limit of infinite momentum cutoff.
The present Hamiltonian renormalization approach is particularly useful when 
several discrete eigenvalues are possible, since it can be diagonalized, in 
a regularized form, in order to obtain physical observables that are well 
defined in the infinite cutoff limit. 

Finally, 
inspired on the Weinberg effort to have a more comprehensible universe,
besides his paradoxical conclusion that 
``The more the universe seems comprehensible, the more it seems pointless!",
let us make more understandable the quantum few-body physics
with fixed-point Hamiltonians.

\section*{Acknowledgements}
This report is dedicated to the memory of Steven Weinberg, as well as 
to celebrate 30 years of his contributions on Nuclear Forces from Chiral
Lagrangians. We thank Alejandro Kievsky for the kind invitation to contribute to this
special volume. This work was partially supported by Funda\c c\~ao de Amparo \`a Pesquisa 
do Estado de S\~ao Paulo (FAPESP) grants 2017/05660-0 (T.F. and L.T.), 2019/10889-1 (V.S.T) and  
2019/00153-8 (M.T.Y.), and by Conselho Nacional de Desenvolvimento Cient\'\i fico 
e Tecnol\'ogico (CNPq) grants 304469/2019-0 (L.T.), 308486/2015-3 (T.F.), 306615/2018-5 (V.S.T.), 
303579/2019-6 (M.T.Y.) and 464898/2014-5 (INCT-FNA).


\begin{thebibliography}{99}

\bibitem{1979wein} S. Weinberg, Phenomenological Lagrangians.  
Physica A {\bf 96}, 327 (1979)

\bibitem{1990wein}  S. Weinberg, Nuclear forces from chiral Lagrangians. 
Phys. Lett. B {\bf 251}, 288 (1990)

\bibitem{1991wein}  S. Weinberg, 
Effective chiral Lagrangians for nucleon-pion interactions and nuclear forces.
Nucl. Phys. B {\bf 363}, 3 (1991)  

\bibitem{1992wein}  S. Weinberg, 
Three-body interactions among nucleons and pions.
Phys. Lett. B {\bf 295}, 114 (1992)

\bibitem{1970wilson}  
K. G. Wilson, Model of coupling-constant renormalization.
Phys. Rev. D{\bf 2}, 1438 (1970

\bibitem{1971wilson} K. G. Wilson, 
Renormalization group and strong interactions. 
Phys. Rev. D {\bf 3}, 1818 (1971)

\bibitem{1974wilson}
K. G. Wilson and J. Kogut, 
The renormalization group and the $\epsilon$ expansion.
Phys. Rep. {\bf 12} {\bf 75}, (1974) 

\bibitem{1983wilson}
K.G. Wilson, The renormalization group and critical phenomena.
Rev. Mod. Phys.{\bf 55},  583 (1983)

\bibitem{1993glazek}  S.D. Glazek and K.G. Wilson, 
Renormalization of Hamiltonians.
Phys. Rev. D{\bf 48}, 5863 (1993)

\bibitem{1993glazek2}
S. Glazek, A. Harindranath, S. Pinsky, J. Shigemitsu, and K. Wilson,
Relativistic bound-state problem in the light-front Yukawa model.
Phys. Rev. D {\bf 47} 1599 (1993)

\bibitem{1994glazek} S. D. Glazek and K. G. Wilson,
Perturbative renormalization group for Hamiltonians.
Phys. Rev. D {\bf 49}, 4214 (1994) 

\bibitem{1997perry}  M.M. Brisudov\'{a}, R.J. Perry and K.G. Wilson, 
Quarkonia in Hamiltonian Light-Front QCD.
Phys.Rev. Lett. {\bf 78},  1227 (1997)

\bibitem{1998glazek} S. D. Glazek and K. G. Wilson
Asymptotic freedom and bound states in Hamiltonian dynamics.
Phys. Rev. D{\bf 57}, 3558 (1998)

\bibitem{1998brodsky} 
S. Brodsky, H.-C. Pauli, S. Pinsky,
Quantum chromodynamics and other field theories on the light cone.
Phys. Rep. {\bf 301}, 299 (1998)

\bibitem{1997lepage}  G. P. Lepage, {\it How to Renormalize the Schr\"{o}dinger
Equations}, Proc. of the VIII Jorge Andr\'{e} Swieca Summer School, pg.135,
World Scientific, Singapore, 1997; nucl-th/9706029

\bibitem{2004glazek} S. D. Glazek and K. G. Wilson,
Universality, marginal operators, and limit cycles.
Phys. Rev. B {\bf 69}, 094304 (2004)

\bibitem{1992amorim}  A.E.A. Amorim, L. Tomio, and T. Frederico, 
Three-boson system with absorptive short range potential.
Phys. Rev. C{\bf 46}, 2224 (1992) 

\bibitem{1997amorim}A.E.A. Amorim, L. Tomio, and T. Frederico,  
Universal aspects of Efimov states and light halo nuclei.
Phys. Rev. C{\bf 56}, 2378 (1997)

\bibitem{1999frederico}
T. Frederico, L. Tomio, A. Delfino, and A.E.A. Amorim, 
Scaling limit of weakly bound triatomic states.
Phys. Rev. A{\bf 60}, R9 (1999)

\bibitem{1999tomio}
L. Tomio, T. Frederico, A. Delfino, and A.E.A. Amorim, 
Three helium atoms and the scaling limit.
Few-Body Syst. Supp. {\bf 10}, 203 (1999)

\bibitem{2000delfino}
A. Delfino, T. Frederico and L. Tomio, 
Low-energy universality in three-body models.
Few-Body Syst. {\bf 28}, 259 (2000) 

\bibitem{2000delf}
A. Delfino, T. Frederico, M.S.Hussein and L. Tomio, 
Virtual states of light non-Borromean halo nuclei.
Phys. Rev. C {\bf 61}, 051301 (2000) 

\bibitem{1999fred}  T. Frederico, V.S. Tim\'{o}teo, and L. Tomio, 
Renormalization of the one-pion-exchange interaction.
Nucl. Phys. A {\bf 653}, 209 (1999)

\bibitem{2000frederico}  T. Frederico, A. Delfino and L. Tomio, 
Renormalization group invariance of quantum mechanics.
Phys. Lett. B {\bf 481}, 143 (2000)

\bibitem{2005tim} V. Tim\'oteo, T. Frederico, A. Delfino, and L. Tomio, 
Recursive renormalization of the singlet one-pion-exchange plus point-like interactions.
Phys. Lett. B {\bf 621}, 109 (2005)

\bibitem{2007tim} V. S. Tim\'oteo, T. Frederico, L. Tomio, and A. Delfino, 
Renomalization of the nn interaction at nnlo: uncoupled peripheral waves. 
Int. Jour. Mod. Phys. E {\bf 16}, 2822 (2007)

\bibitem{2007timNPA} V. Tim\'oteo, T. Frederico, A. Delfino, and L. Tomio, 
Subtractive renormalization of the next-to-leading order NN interaction.
Nucl. Phys. A {\bf 790}, 406c (2007)

\bibitem{2011tim} V. Tim\'oteo, T. Frederico, A. Delfino, and L. Tomio, 
Nucleon-nucleon scattering within a multiple subtractive renormalization approach.
Phys. Rev. C {\bf 83}, 064005 (2011)

\bibitem{2011std} S. Szpigel, V. S. Tim\'oteo and F. O. Dur\~aes, 
Similarity renormalization group evolution of chiral effective nucleon–nucleon potentials 
in the subtracted kernel method approach.
Annals of Physics {\bf 326}, 364 (2011)

\bibitem{2012szpigel} S. Szpigel and V. S. Tim\'oteo, 
Power counting and renormalization group invariance in the subtracted kernel 
method for the two-nucleon system.
J. Phy. G: Nuclear and Particle Physics, {\bf 39}, 105102 (2012)

\bibitem{2011birse} M. C. Birse, 
The renormalization group and nuclear forces. 
Phil. Trans. R. Soc. A {\bf 369}, 2662 (2011)

\bibitem{2012fred} T. Frederico, A. Delfino, L. Tomio and M. T. Yamashita,
Universal aspects of light halo nuclei.
Prog. Part. Nucl. Phys. {\bf 67}, 939 (2012)

\bibitem{2017batista} E.F. Batista, S. Szpigel, V.S. Tim\'oteo, 
Renormalization of Chiral Nuclear Forces with Multiple Subtractions in Peripheral Channels.
Adv. High Energy Phys. {\bf 2017}, 2316247 (2017)

\bibitem{2020ham} H.-W. Hammer, S. K\"onig, and U. van Kolck, 
Nuclear effective field theory: Status and perspectives.
Rev. Mod. Phys. {\bf 92}, 025004 (2020)

\bibitem{2020epel} E. Epelbaum, A. M. Gasparyan, J. Gegelia, Ulf-G. Meißner, X.-L. Ren, 
How to renormalize integral equations with singular potentials
in effective field theory.
Eur. Phys. J. A {\bf 56}, 152 (2020)

\bibitem{2021batista} E. F. Batista, S. Szpigel, V. S. Tim\'oteo, 
Pions and Contacts at N4LO: Some details on the chiral nuclear force.
Ann. of Phys. {\bf 425}, 168383 (2021)

\bibitem{2021entem} D. R. Entem and J. A. Oller, 
Non-perturbative methods for NN singular interactions.
Eur. Phys. J. Spec. Top. {\bf 230}, 1675 (2021)

\bibitem{2021timoteo} V. S. Tim\'oteo,
Computational approaches for three-nucleon systems.
Ann. of Phys. {\bf 432}, 168573 (2021)

\bibitem{2001frederico} T. Frederico and H. C. Pauli, 
Renormalization of an effective light-cone QCD-inspired theory for the 
pion and other mesons. 
Phys. Rev. D {\bf 64} 054007 (2001)

\bibitem{1995Adhikari1}  S. K. Adhikari, T. Frederico and I. D. Goldman, 
Perturbative Renormalization in Quantum Few-Body Problems.
Phys. Rev. Lett. {\bf 74}, 487 (1995)

\bibitem{1995Adhikari2} S. K. Adhikari and T. Frederico, 
Renormalization Group in Potential Scattering.
Phys. Rev. Lett. {\bf 74}, 4572 (1995)

\bibitem{wein1}  S. Weinberg, ``The Quantum Theory of Fields Vol. I,
Foundations'', Cambridge University Press 1995; and
``The Quantum Theory of Fields Vol. II, Modern Applications'', 
Cambridge University Press, 1996

\bibitem{1998fisher}  M. E. Fisher, 
Renormalization group theory: Its basis and formulation in statistical physics.
Rev. Mod. Phys. {\bf 70}, 653 (1998)

\bibitem{2002zinnjustin} J. Zinn-Justin, Quantum Field Theory and Critical
Phenomena, 4th ed., Claredon Press-Oxford 2002

\bibitem{1970Callan} C. G. Callan, 
Broken Scale Invariance in Scalar Field Theory.
Phys. Rev. D{\bf 2}, 1541 (1970)

\bibitem{1970Symanzik1} K. Symanzik, 
Renormalizable models with simple symmetry breaking.
Comm. Math. Phys. {\bf 16}, 48 (1970)

\bibitem{1970Symanzik2} K. Symanzik, 
Small distance behaviour in field theory and power counting.
Comm. Math. Phys. {\bf 18}, 227 (1970)

\bibitem{1993georgi} H. Georgi, 
Effective field theory.
Ann. Rev. Nucl. Part. Sc. {\bf 43}, 209 (1993)

\bibitem{preprint} T. Frederico, A. Delfino, L. Tomio, and 
V.S. Tim\'oteo, Fixed-Point Hamiltonians in Quantum Mechanics.
	arXiv:hep-ph/0101065 (2001)
	
\bibitem{stapp} H. P. Stapp, T. J. Ypsilantis, and 
N. Metropolis, 
Phase-shift analysis of 310-Mev proton-proton scattering experiments.
Phys. Rev. {\bf 105}, 302 (1957)

\bibitem{2008bira} B. Long and U. van Kolck,
Renormalization of singular potentials and power counting.
Ann. of Phys. {\bf 323}, 1304 (2008)

\bibitem{2005Nogga} A. Nogga, R. G. E. Timmermans, and U. van Kolck,
Renormalization of one-pion exchange and power counting,
Phys. Rev. C {\bf 72}, 054006 (2005)

\bibitem{nij} V. G. J. Stocks, R. A. M. Klomp, C. P. F. Terheggen, 
and J.J. de Swart, Construction of high-quality NN potential models.
Phys. Rev. C {\bf 49}, 2950 (1994)	

\bibitem{YamPLB2008b} M. T. Yamashita, T. Frederico, L. Tomio,
Neutron–$^{19}$C scattering near an Efimov state. 
Phys. Lett. B {\bf 670}, 49 (2008)

\bibitem{2021madeira}
L. Madeira, T. Frederico, S. Gandolfi, L. Tomio, and M. T. Yamashita,
Quantum Monte Carlo studies with microscopic two- and three-body interactions of a trimer 
scaling function.
Preprint arXiv:2106.09058 (2021)

\bibitem{HammerJPG2017} H.-W. Hammer, C. Ji, D. R. Phillips, 
Effective field theory description of halo nuclei. J. Phys. G {\bf 44}, 10 (2017)

\bibitem{AfnPRC2004} I. R. Afnan and D. R. Phillips,  
Three-body problem with short-range forces: 
Renormalized equations and regulator-independent results. 
Phys. Rev. C  {\bf 69}, 034010 (2004)

\bibitem{ShalchiPLB2017} M. A. Shalchi, M. T. Yamashita, M. R. Hadizadeh, T. Frederico, L. Tomio, 
Neutron$-^{19}$C scattering: Emergence of universal properties in a finite 
range potential. Phys. Lett. B {\bf 764}, 196 (2017)

\bibitem{Efimov}   V.  Efimov,   Energy  levels  arising  from  resonant  two-body 
forces in a three-body system. Phys. Lett. B {\bf 33}, 563 (1970)

\bibitem{EfimovNPA1981} V.  Efimov, Qualitative treatment of three-nucleon properties, 
Nucl. Phys. A, {\bf 362}, 45 (1981)

\bibitem{ShalchiPRA2018} 
M. A. Shalchi, M. T. Yamashita, M. R. Hadizadeh, E. Garrido, L. Tomio, and 
T. Frederico, Probing Efimov discrete scaling in an atom-molecule collision.
Phys. Rev.  A {\bf 97}, 012701  (2018)

\bibitem{ShalchiPRA2020} 
M. A. Shalchi, M. T. Yamashita, T. Frederico, L. Tomio, Cold atom-dimer 
reaction rates with $^4$He, $^{6,7}$Li, and $^{23}$Na.  
Phys. Rev. A {\bf 102}, 062814 (2020)

\bibitem{YamEPL2006} M. T. Yamashita, L. Tomio, A. Delfino, 
T. Frederico, Four-boson scale near a Feshbach resonance.
EPL {\bf 75}, 555 (2006)

\bibitem{HadPRL2011} 
M. R. Hadizadeh, M. T. Yamashita, L. Tomio, A. Delfino, T. Frederico,
Scaling properties of universal tetramers.
Phys. Rev. Lett. {\bf 107}, 135304 (2011)

 \bibitem{Skorniakov1957} G. V. Skorniakov and K. A. Ter-Martirosian,  Three body problem for 
 short range forces. I. Scattering of low en-ergy neutrons by deuterons.  
 Soviet Phys. JETP {\bf 4}, 648 (1957)

\bibitem{ThomasPR1935} L. H. Thomas, The interaction between a neutron and a proton  and  the  
structure  of  H$^3$. Phys.  Rev. {\bf 47}, 903 (1935)

\bibitem{FredFBS2011}
T. Frederico, L. Tomio, A. Delfino, M. R. Hadizadeh, M. T. Yamashita,  
Scales and universality in few-body systems.
Few-Body Syst. {\bf 51}, 87 (2011)

\bibitem{HadPRA2012} M. R. Hadizadeh, M. T. Yamashita, L. Tomio, 
A. Delfino, T. Frederico, 
Binding and structure of tetramers in the scaling limit.
Phys. Rev. A {\bf 85}, 023610 (2012)

\bibitem{FredFBS2012} T. Frederico, A. Delfino, M. R. Hadizadeh, L. Tomio, 
and M. T. Yamashita, 
Universality in four-boson systems.
Few-Body Syst. {\bf 54}, 559 (2012)

\bibitem{TomioFBS2014}
L. Tomio, M. R. Hadizadeh, M. T. Yamashita, T. Frederico, A. Delfino, 
Trimer-tetramer interwoven states in the scaling limit. 
Few-Body Syst. {\bf 55}, 949 (2014)

\bibitem{DeltuvaFBS2011} A. Deltuva, R. Lazauskas, and L. Platter,
Universality in four-body scattering.
Few-Body Syst. {\bf 51}, 235 (2011)

\bibitem{ZinnerJPG2013} 
N. T. Zinner and A. S. Jensen, 
Comparing and contrasting nuclei and cold atomic gases.
J. Phys. G {\bf 40}, 053101 (2013)

\bibitem{NaidonRPP2017} P. Naidon and S. Endo,  
Efimov Physics: a review. 
Rept. Prog. Phys. {\bf 80}, 056001 (2017)

\bibitem{GreeneRMP2017} C. H. Greene, P. Giannakeas, and J. Perez-Rios, 
Universal few-body physics and cluster formation.  
Rev. Mod. Phys. {\bf 89}, 035006 (2017)

\bibitem{FredFBS2019} T. Frederico, W. de Paula, A. Delfino, and M. T. Yamashita, 
L. Tomio, 
Four-boson continuous scale symmetry breaking.
Few-Body Syst. {\bf 60}, 46 (2019)

\bibitem{dePaulaJPG2020} W. de Paula, A. Delfino, T. Frederico, and L. Tomio, 
Limit cycles in the spectra of mass imbalanced many-boson system.
J. Phys. B  {\bf 53},  205301 (2020)

\bibitem{BazakPRL2019} B. Bazak, J. Kirscher, S. K\"onig, M. P. Valderrama, 
N. Barnea, and U. van Kolck,
Four-body scale in universal few-boson systems.
Phys. Rev. Lett. {\bf 122}, 143001 (2019)

\bibitem{GattobigioPRA2014} M. Gattobigio, A. Kievsky, 
Universality and scaling in the N-body sector of Efimov physics. 
Phys. Rev. A {\bf 90}, 012502 (2014)

\bibitem{KievskyPRA2014} A. Kievsky, N. K. Timofeyuk, M. Gattobigio, N-boson spectrum 
from a discrete scale invariance. 
Phys. Rev. A {\bf 90}, 032504 (2014)

\bibitem{birse} M. C. Birse, J. A. McGovern, and K. G. Richardson, 
A renormalisation-group treatment of two-body scattering.
Phys. Lett. B {\bf 464}, 169 (1999)

\end{thebibliography}
\end{document}